\documentstyle[12pt]{article}
\newcommand{\be}[1]{\begin{equation} \label{(#1)}}
\newcommand{\ee}{\end{equation}}
\newcommand{\ba}[1]{\begin{eqnarray} \label{(#1)}}
\newcommand{\ea}{\end{eqnarray}}
\newcommand{\nn}{\nonumber}
\newcommand{\rf}[1]{(\ref{(#1)})}

\def\pmb#1{\setbox0=\hbox{#1}%
  \kern-.015em\copy0\kern-\wd0
  \kern.03em\copy0\kern-\wd0
  \kern-.015em\raise.0233em\box0 }

\begin{document}

\begin{center}
{\bf B-L-violating Masses in Softly Broken Supersymmetry.}
\bigskip

{M. Hirsch, H.V. Klapdor-Kleingrothaus and S.G. Kovalenko$^*$
\bigskip

{\it
Max-Planck-Institut f\"{u}r Kernphysik, P.O. 10 39 80, D-69029,
Heidelberg, Germany}

\bigskip

$^*${\it Joint Institute for Nuclear Research, Dubna, Russia}
}
\end{center}

\begin{abstract}
We prove a general low-energy theorem establishing a generic relation
between the neutrino Majorana mass and the superpartner sneutrino 
B-L-violating "Majorana"-like mass term. 
The theorem states that, if one of these two quantities is non-zero 
the other one is also non-zero and, vice versa, if one of them vanishes 
the other vanishes, too. The theorem is a consequence of the underlying 
supersymmetry (SUSY) and valid for any realistic gauge model 
with weak scale softly broken SUSY. 
\end{abstract}
Neutrinos are believed to be massive particles. Despite the lack 
of unambiguous experimental confirmation of this belief there are 
insisting indications for non-zero neutrino masses from  cosmology,
the solar and atmospheric neutrino puzzles 
(for recent review see \cite{Smirnov})
as well as from recent LSND results on possible 
$\bar\nu_e-\bar\nu_{\mu}$ neutrino oscillations \cite{LSND}.    

Among the known explanations for the extreme smallness of the neutrino 
mass compared to masses of the other fermions the most natural one 
is based on the see-saw mechanism \cite{see-saw}. 
It leads to a B-L-violating 
Majorana mass term for the neutrino.  Various 1-loop contributions to the
neutrino self-energy, widely discussed in the literature
\cite{Lee84}-\cite{Zee}, also induce a small Majorana mass for neutrinos. 
 Furthermore, the Grand Unification paradigm definitely prefers a
Majorana mass for neutrinos. 
Due to these arguments, it has become a common trend to think of
neutrinos as Majorana particles.  

In supersymmetric (SUSY) models the neutrino $\nu$ has its scalar 
superpartner 
the sneutrino $\tilde\nu$. Given that they are components of 
the same superfield
one may suspect a certain interplay between the neutrino and sneutrino 
properties at low energies as a relic of the underlying supersymmetry.  

In the present note we prove a low-energy theorem establishing 
an intimate relation between the neutrino Majorana mass term 
and the B-L-violating as well as B-L-conserving sneutrino mass terms.
Our consideration refers to the general structure of the low-energy 
effective Lagrangian assuming weak scale softly broken
supersymmetry and stability of the ground state after 
electro-weak symmetry breaking.
The proof of the low-energy theorem, consisting of three
statements, is based on symmetry arguments and is of a general significance.   
 
The effective Lagrangian of a generic model of  weak scale supersymmetry
contains after electro-weak symmetry breaking the following terms
\ba{L1}
{\cal L} &=& - \sqrt{2} g \epsilon_i \cdot 
\overline{\nu}_L\chi_i\tilde\nu_L   
- g \epsilon_i^- \cdot \overline{e}_L\chi_i^-\tilde\nu_L   
- g \epsilon_i^+ \cdot \overline{\nu}_L\chi_i^+\tilde e_L + \\ \nn
&+&  \frac{g}{\sqrt{2}}\cdot \overline{\nu}_L \gamma^{\mu} e_L W^+_{\mu}
+ g \cdot \bar\chi_i\gamma^{\mu} (O^L_{ij}P_L +
                                  O^R_{ij}P_R)\chi^+_j W^-_{\mu} +  
                                                             ... + h.c. 
\ea
Dots denote other terms which are not essential 
for further consideration.  
Here, $\tilde\nu_L$ and $\tilde e_L$
represent scalar superpartners of the left-handed neutrino $\nu_L$ and
electron $e_L$ fields. The chargino $\chi^{\pm}_i$ and neutralino
$\chi_i$ are superpositions of the gaugino and the higgsino fields.
The contents of these superpositions depends on the model. 
Note that the neutralino is a Majorana field $\chi_i^c =  \chi_i$.
The explicit form of the coefficients $\epsilon, \ \epsilon^{\pm}_i $ and
$O^{L,R}_{ij} $ is also  unessential.  For the case of the MSSM one
can  find them, for instance in \cite{Haber}. 
Eq. \rf{L1} is a general consequence of the underlying
weak scale supersymmetry (softly broken) and the spontaneously 
broken electro-weak gauge symmetry. 

Now let us assume the neutrino acquires a Majorana mass 
\ba{nu_M}
{\cal L}^{\nu}_M = -\frac{1}{2} (m_M^{\nu} \overline{\nu^c} \nu + h.c.),  
\ea 
where $\nu = \nu^c$ is a Majorana field. 
The further proof does not depend on the mechanism generating this mass
term in the low-energy Lagrangian.
For the sake of simplicity and  without any loss of generality we
ignore possible neutrino mixing.  

Prove the following statements.
 
\underline{Statement 1}: If $m_M^{\nu}\neq 0$ then in the low-energy
Lagrangian also the sneutrino "Majorana"-like
B-L-violating mass term is present
\ba{snu_M}
{\cal L}^{\tilde\nu}_M  = -\frac{1}{2} (\tilde m_M^ 2 
\tilde\nu_L \tilde\nu_L + h.c.)
\ea
with $\tilde m_M^2\neq 0$. Note that $\tilde m_M^2$ is not a
positively defined parameter. 

\underline{Statement 2} is  inverse to the statement 1: 
If $\tilde m_M^2\neq 0$ in Eq. \rf{snu_M}, then in Eq. \rf{nu_M}  
$m_M^{\nu}\neq 0$. 

First notice that in the presence of 
a non-zero Majorana neutrino mass term in Eq. \rf{nu_M} the
"Majorana"-like  
sneutrino mass term in Eq. \rf{snu_M} is generated at
the 1-loop level as shown in Fig.1(a) with neutrino and neutralino
internal lines. An opposite statement based on the
1-loop diagram in Fig.1(b) is also true. The  B-L-violating sneutrino
propagator in Fig.1(b) is proportional to $\tilde m_M^2$ and explicitly
derived below. Here we do not need
detailed calculations and write down schematically
\ba{1-loop}
\tilde m_M^2 &=
& \frac{1}{4\pi}(g^2 + g'^2) m_M^{\nu}  M_{\chi_i} \gamma_i  + 
A_{\tilde\nu},\\
 \label{(1-loop2)}
m_M^{\nu} &=& \frac{1}{4\pi}(g^2 + g'^2)   
\frac{\tilde m_M^2}{\tilde m_D^2} M_{\chi_i} \beta_i + A_{\nu}. 
\ea
Here $m_D$ is a mass parameter explained below, $g$ and $g'$ are 
the $SU(2)_L\times U_{1Y}$ gauge coupling
constants, $M_{\chi_i}$ are the neutralino masses and $\beta_i, \gamma_i$
are functions depending on neutralino mixing coefficients and on 
the masses of particles in the loop. $ A_{\tilde\nu} $ and $A_{\nu}$ 
represent any other possible contributions. 
The explicit form of $A_i, \beta_i,
\gamma_i$ is not essential for us.  Important is just the presence of 
a correlation between $\tilde m_M^2$ and $m_M^{\nu}$ which 
we write down in
general as 
$
\tilde m_M^2 =  f(m_M^{\nu}),\  m_M^{\nu} = \phi(\tilde m_M^2).
$
Now we are going to prove that 
\ba{funct_0}
\tilde m_M^2 =  f(m_M^{\nu}=0) = 0, \ \ \ \ 
       m_M^{\nu} = \phi(\tilde m_M^2=0) = 0.
\ea
One can expect such properties of the functions $f$ and $\phi$ from 
Eqs. \rf{1-loop}-\rf{1-loop2}. Indeed, $\tilde m_M^2 = 0$ in the
left-hand side of Eq. \rf{1-loop} strongly disfavors $m_M^{\nu}\neq 0$.
Similarly, $m_M^{\nu}= 0$ in the left-hand side of Eq. \rf{1-loop2} 
strongly disfavors $\tilde m_M^2\neq 0$. This is because vanishing
left-hand sides of Eqs. \rf{1-loop}-\rf{1-loop2} requires either 
vanishing of both terms in the right-hand sides or their 
net cancelation. The latter is unlikely since it implies unnatural
fine-tuning of certain parameters. More serious, it is unstable under 
radiative corrections. Even if the fine-tuning was done by hand 
it would be spoiled in higher orders of perturbation theory. 
To guarantee the cancelation of both
terms in the right-hand sides of Eqs. \rf{1-loop}-\rf{1-loop2} 
in all orders of perturbation theory one needs a special
unbroken symmetry. 
The Lagrangian \rf{L1} does not posses any
continuous symmetry having non-trivial B-L transformation properties.
However, there might be an appropriate discrete symmetry. 

Let us specify this discrete symmetry group by the following 
field transformations
\ba{discrete}
\nu &\rightarrow & \eta_{\nu} \nu, \ \ \
\tilde\nu\rightarrow\eta_{\tilde\nu} \tilde\nu, \ \ \
e_L \rightarrow  \eta_{e} e_L, \ \ \ 
\tilde e_L  \rightarrow \eta_{\tilde e} \tilde e_L, \\ \nn 
W^+ & \rightarrow & \eta_{_W} W^+,  \ \ \ 
\chi_{i}\rightarrow\eta_{\chi_{i}} \chi_{i},\ \ \ 
\chi^+  \rightarrow  \eta_{\chi^+} \chi^+.
\ea
Here $\eta_i$ are phase factors.  
Since the Lagrangian \rf{L1} is assumed to be invariant under these
transformations one obtains the following relations
\ba{symm_rel}
\eta_{\nu}^* \eta_{\tilde\nu} \eta_{\chi_i} &=& 1, \ \ \  
\eta_{e} \eta_{\chi^+}\eta_{\tilde\nu}^* = 1, \\ \nn
\eta_{e} \eta_{_W} \eta_{\nu}^* &=& 1, \ \ \ 
\eta_W^* \eta_{\chi^+} \eta_{\chi_i}^* = 1, \ \ \ ...  
\ea
Dots denote other relations which are not essential here.
The complete set of these equations defines the admissible 
discrete symmetry group of the Lagrangian in Eq. \rf{L1}.

Solving these equations, one finds
\ba{sol}
\eta_{\nu}^2 = \eta_{\tilde\nu}^2.
\ea
This relation proves the statements 1,2 and the corresponding properties
expressed by Eqs. \rf{funct_0}. To see this we note that 
the B-L-violating mass terms in Eq. \rf{nu_M} or in Eq. \rf{snu_M} 
are forbidden by this symmetry if $\eta_{\nu}^2\neq 1$ or 
$\eta_{\tilde\nu}^2\neq 1$. Contrary, if  
$\eta_{\nu}^2= 1$, this mass term is not protected by the symmetry 
and appears in higher orders of perturbation theory, even if 
it does not exist at the tree-level.  Relation \rf{sol}
claims that if the neutrino Majorana mass term in Eq. \rf{nu_M} 
is forbidden, {\it i.e.} $m_M^{\nu} = 0$, then the sneutrino 
Majorana-like mass
term in Eq. \rf{snu_M} is also forbidden, $\tilde m_M = 0$, and vice 
versa. If one of them is not forbidden then both are not forbidden.
Thus, statements 1 and 2 as well as the explicit relations in 
Eqs. \rf{funct_0}
are proven.

One can derive the following corollary from statements 1,2.

{\it Corollary}: 
 If one of the two B-L-violating masses, either $m_M^{\nu} $ or 
$\tilde m_M^2$, vanishes, then the other one vanishes too.

Let us turn to the last statement.
 
\underline{Statement 3}: In the presence of $\tilde m_M^2\neq 0$
in Eq. \rf{snu_M} there must exist a "Dirac"-like B-L-conserving
sneutrino  mass term
\ba{snu_D}
{\cal L}^{\tilde\nu}_D  = - \tilde m_D^2 \tilde\nu_L^* \tilde\nu_L
\ea
with $\tilde m_D^2\geq|\tilde m_M^2|$.

To prove this statement consider the combined sneutrino mass term 
${\cal L}^{\tilde\nu}_{mass} = {\cal L}^{\tilde\nu}_M + 
{\cal L}^{\tilde\nu}_D$ 
and use the real field representation for the  complex 
scalar sneutrino field
$\tilde\nu = (\tilde\nu_1 + i \tilde\nu_2)/\sqrt{2}$, where 
$\tilde\nu_{1,2}$ are real fields. Then
\ba{snu_DM}
{\cal L}^{\tilde\nu}_{mass} = - \frac{1}{2}( \tilde m_M^2
\tilde\nu_L \tilde\nu_L + h.c.) - \tilde m_D^2 \tilde\nu_L^* \tilde\nu_L
 = - \frac{1}{2}  \tilde m_1^2
\tilde\nu_1^2 -  \frac{1}{2}  \tilde m_2^2 \tilde\nu_2^2 
\ea
where $\tilde m_{1,2}^2 = \tilde m_D^2 \pm |\tilde m_M^2|$. Assume the
vacuum state is stable. Then $\tilde m_{1,2}^2\geq 0$ or 
$\tilde m_D^2 \geq |\tilde m_M^2|$, otherwise vacuum is unstable and
the subsequent spontaneous symmetry breaking occurs via 
non-zero vacuum expectation
values of the sneutrino fiels $<\tilde\nu_i>\neq 0$. 
The broken symmetry in this case is the R-parity. 
It is a discrete symmetry  defined as   
$R_p = (-1)^{3B+L+2S}$, where $S,\ B$ and $L$
are the spin, the baryon and the lepton quantum number. 

This completes the proof of the theorem consisting of the above 
three statements.

From the above consideration it follows that a self-consistent
structure of mass terms of the neutrino-sneutrino sector is 
\ba{complete}
{\cal L}^{\nu\tilde\nu}_{mass} = -\frac{1}{2} (m_M^{\nu}
\overline{\nu^c} \nu + h.c) - \frac{1}{2}( \tilde m_M^2
\tilde\nu_L \tilde\nu_L + h.c.) - 
\tilde m_D^2 \tilde\nu_L^* \tilde\nu_L.  
\ea 
The Dirac neutrino mass term 
$m_D^{\nu} (\bar\nu_L \nu_R + \bar\nu_R\nu_L)$ can also be introduced 
but it is  not required by the self-consistency arguments.
  
It is instructive to derive an explicit form of the above mentioned 
B-L-violating sneutrino propagator. It can be done by the use of 
the real field representation as in Eq. \rf{snu_DM}. 
Let us consider for comparison both the B-L-conserving 
$\Delta_{\tilde\nu}^D $ 
and the B-L-violating $\Delta_{\tilde\nu}^M$  sneutrino propagators 
\begin{eqnarray} \label{propagators_1}
&&\Delta_{\tilde\nu}^D(x-y) =
<0|T(\tilde\nu(x)\tilde\nu^*(y))|0> =\\  \nonumber
&&= \frac{1}{2}<0|T(\tilde\nu_1(x)\tilde\nu_1(y))|0>\  + \ 
\frac{1}{2}<0|T(\tilde\nu_2(x)\tilde\nu_2(y))|0> = \\ \nonumber
&&=-\frac{i}{2}(\Delta_{\tilde m_1}(x-y) + 
\Delta_{\tilde m_2 }(x-y)),  \\
&&\Delta_{\tilde\nu}^M(x-y)= <0|T(\tilde\nu(x)\tilde\nu(y))|0> =  
\\ \nonumber
&&= \frac{1}{2}<0|T(\tilde\nu_1(x)\tilde\nu_1(y))|0> \  -  \  
 \frac{1}{2} <0|T(\tilde\nu_2(x)\tilde\nu_2(y))|0> = \\ \nonumber
&&= -\frac{i}{2}(\Delta_{\tilde m_1}(x-y) - \Delta_{\tilde m_2 }(x-y)),  
\end{eqnarray}
where
\begin{eqnarray} \label{def_1}
\Delta_{\tilde m_i}(x) = \int\frac{d^4 k}{(2\pi)^4}
\frac{e^{-ikx}}{\tilde m_i^2 - k^2 - i\epsilon}
\end{eqnarray}
is the ordinary propagator for a scalar particle with mass $\tilde m_i$. 
Using the definition of $\tilde m_{1,2}$ as in Eq. \rf{snu_DM} 
one finds    
\begin{eqnarray} \label{propagators_2} 
\Delta_{\tilde\nu}^D(x) &=& \int\frac{d^4 k}{(2\pi)^4}
\frac{\tilde m_D^2 - k^2}{(\tilde m_1^2 - k^2 - i\epsilon)
(\tilde m_2^2 - k^2 - i\epsilon)} e^{-ikx}, \\
\Delta_{\tilde\nu}^M(x) &=& -  \tilde m_M^2 \int\frac{d^4 k}{(2\pi)^4}
\frac{e^{-ikx}}{(\tilde m_1^2 - k^2 - i\epsilon)
(\tilde m_2^2 - k^2 - i\epsilon)}. 
\end{eqnarray}
It is seen that in absence of the B-L-violating sneutrino 
Majorana-like mass term 
$\tilde m_M^2 =0$ the B-L-violating propagator vanishes while 
the B-L-conserving one becomes the ordinary propagator of 
a scalar particle
with mass $\tilde m_{1} = \tilde m_{2} = \tilde m_D$. 

In presence of the B-L-violating sneutrino Majorana-like mass term 
the complex scalar sneutrino field splits into two real mass
eigenstate fields $\tilde\nu_{1,2}$  with different masses 
$\tilde m_{1,2}$. The square mass splitting is $2 \tilde m_M^2$.

This sneutrino mass splitting parameter can be probed by searching for
B-L-violating exotic processes such as neutrinoless double beta
decay. It is obvious from Eq. \rf{1-loop}-\rf{1-loop2} that certain
constraints on $\tilde m_M^2 $ can be also obtained
from the experimental upper bound on the neutrino mass. 
We are going to analyze these constraints in a separate paper.

In summary, we have proven a low-energy theorem for weak scale softly
broken supersymmetry relating the B-L-violating mass terms of the
neutrino and the sneutrino.

\bigskip
\centerline{\bf ACKNOWLEDGMENTS}

We thank V.A.~Bednyakov,  for helpful discussions.
The research described in this publication was made 
possible in part (S.G.K.) by
Grant GNTP 315NUCLON from the Russian ministry of science. 
M.H. would like to thank the Deutsche Forschungsgemeinschaft
for financial support by grants kl 253/8-2 and 446 JAP-113/101/0.

{\large\bf Figure Captions}\\

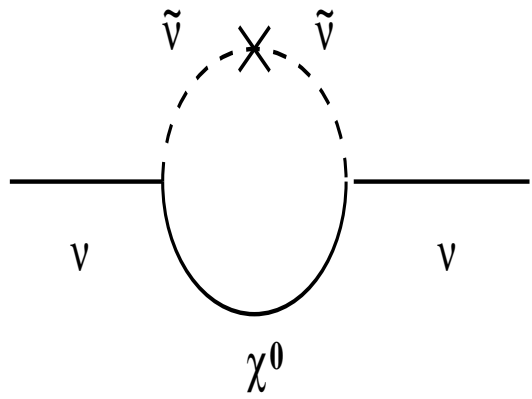
Fig.1.       1-loop contributions to 
            (a) the neutrino Majorana mass $m_M^{\nu}$ and 
            (b) the sneutrino B-L-violating mass $\tilde{m}^2_M$.  
              Crossed (s)neutrino lines correspond to 
               the B-L-violating propagators.

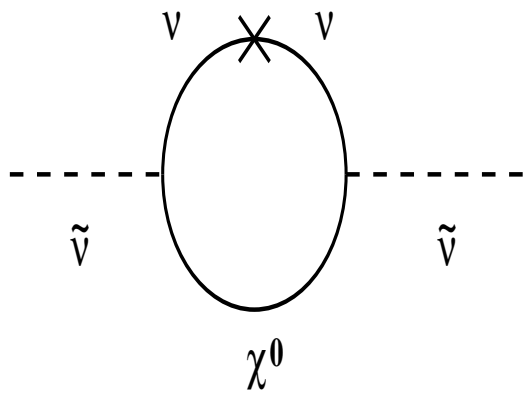
\end{document}